\documentclass[10pt]{article}
\usepackage{epsfig}
\begin{document}
\renewcommand{\thefootnote}{\fnsymbol{footnote}}

	\begin{center}
   {\Large \bf Intergalactic cold dust in the NGC\,4631 group}
\vspace{3em}

   N. Neininger$^{\ast\dag}$\footnote[3]{To whom reprint requests
should be addressed. e-mail: nneini@astro.uni-bonn.de}
           and
   M. Dumke$^{\dag\S}$

\vspace{1em}           
{\small
   $^{\ast}$Radioastronomisches Institut der Universit\"at Bonn,
              Auf dem H\"ugel 71, 
              D-53121 Bonn, Germany; 
   $^{\dag}$Institut de Radioastronomie Millim\'etrique,
              300, rue de la piscine,
              F-38406 St.Martin d'H\`eres, France; 
   $^{\S}$Max-Planck-Institut f\"ur Radioastronomie,
              Auf dem H\"ugel 69,
              D-53121 Bonn, Germany 
}
          \end{center}

\noindent
{\small \it Edited by Marshall H. Cohen, California Institute of
Technology, Pasadena, CA, and approved April 2, 1999 (received for
review March 5, 1999)}

\subsection*{Abstract}

We have detected extraplanar cold dust at distances out to $> 10\,$kpc,
situated in the halo of the interacting galaxy NGC\,4631.  The dust
emission disk is much thinner than the warped H{\sc i} disk and new
structures emerge.  In particular, a giant arc has been found that is
linked to anomalies in the kinematical structure of the atomic
gas. Most of the extraplanar dust is closely associated with H{\sc i}
spurs that have been found earlier [Weliachew, L., Sancisi, R. \&
Gu\'elin, M. (1978) {\it Astron. Astrophys.} 65, 37-45; Rand,
R.J. (1994) {\it Astron. Astrophys.} 285, 833-856]. These spurs
obviously are traces of the interaction [Combes, F. (1978) {\it
Astron. Astrophys.} 65, 47-55].  The dust emission within the plane
reaches the border of the optical disk.

The activity of the disk of NGC\,4631 is moderately enhanced by the
interaction, but no gas moving in the $z$-direction could be found
[Rand, R.J., Kulkarni, S.R. \& Hester, J.J. (1992) {\it Astrophys. J.}
396, 97-103; Golla, G., Dettmar, R.-J. \& Domg\"orgen, H. (1996) {\it
Astron. Astrophys.} 313, 439-447].  Hence, it seems unlikely that
strong winds have deposited the high-$z$ dust.  Instead, the
coincidence with the H{\sc i} features suggests that we see a track
left behind by the interaction.  In addition, the H{\sc i} shows a
supershell formed by an impact [Rand, R.J. \& Stone, J.M. (1996) {\it
Astron. J.} 111, 190-196] in the zone where the dust trail crosses the
disk.  This region is also characterized by disturbances in the
distribution of the H$\alpha$ light.  The masses associated with the
dust can be estimated only very roughly on the basis of the existing
data; they are of the order of a few $10^9$ M$_{\odot}$ of gas.

\section{Introduction}

Cold dust has come into focus only recently because it had to await
the development of sensitive millimeter/sub-millimeter bolometer
arrays to be detectable unambiguously.  The IRAS survey could provide
only hints at its existence because it was blinded by the strong
emission from the small percentage of warmer dust that is radiating
far more brightly.  The large amount of cold dust ($T_{d}\le 25$ K)
can be detected only at (sub)mm wavelengths where the radiation of the
warmer components has vanished.  To give an example: The peak
brightness of a blackbody at 30\,K is 30$\times$ higher than that of a
15-K object, all other parameters being equal.  On the other hand, the
radiation of a blackbody at 15\,K peaks at $\sim200\,\mu$m and remains
more than one order of magnitude brighter at $\lambda$ 1.2\,mm than
that of a blackbody at 30\,K with the same peak brightness.  Now, the
emissivity of interstellar dust is roughly proportional to $T_{d}^{6}$
-- for a blackbody $B(T)=\sigma\,T^4$ -- so even a very large amount
of cold dust emits only weakly.  Because of this $T^{6}$ dependence of
the emission, a very large energy input is needed to heat dust, and
the majority of it remains at lower temperatures.  This cold component
thus is an important tracer, and, indeed, it may represent $> 90$\% of
the interstellar dust (cf.\ refs.\ 1 and 2).

In itself, the contribution of the dust to the total mass of a galaxy
is $< 1$\% of the gas mass, but there are indications that the
dust-to-gas ratio is relatively constant, independent of the type of
gas (atomic or molecular).  Indeed, the dust grains are believed to
play a crucial r\^ole in the formation of molecules.  The study of
their properties thus also offers an independent means of studying the
molecular gas content of galaxies.  This is important because the
standard practice of observing the CO molecule and deriving, thereby,
the properties of the H$_{2}$ has substantial uncertainties
particularly concerning the derived masses.  On the other hand,
investigating the cold dust is technically difficult and cannot
provide any information about the kinematics because it is based on
broadband continuum observations.

In several runs, the IRAM (Institute for Radio Astronomy in the
Millimeter domain) 30-m telescope equipped with MPIfR
(Max-Planck-Institute for Radioastronomy) bolometer arrays has been
used to map nearby galaxies in the $\lambda$ 1.2-mm continuum
emission.  The first maps led to the impression that it is well
correlated with the CO emission, and drops off similarly steeply with
increasing distance from the center.  This behavior was shown, for
example, for galaxies NGC\,891 (3), M\,51 (4), and NGC\,4631 (5).  It
soon became evident, however, that this is not generally the case.
The Sb galaxy NGC\,4565 is significantly more extended in the emission
of $\lambda$ 1.2-mm continuum than in that of the CO line (6).  The
cold dust is even detected in the warped outermost rim of the disk.
As an intermediate case, NGC\,5907 also shows an extended dust disk
(7).  The sensitivity needed to detect this extended emission has been
achieved only recently, however, and the sample is still small.  So it
is not yet clear what determines the extent of the cold dust -- the
profile of the cold dust along the major axis in NGC\,891 remains very
close to the rapidly vanishing CO, even when studied with much higher
sensitivity than previously published (R.  Zylka, personal
communication).

\section{The observations and the object: NGC 4631}
\subsection{Observational details}

All recent maps were obtained with bolometer arrays consisting of 19
elements whose sensitivity is in practice about a factor of 2 better
than that of the 7-element detector used before.  The 19-element
bolometer array has a bandwidth of $\sim 80$\,GHz centered at $\sim$
230\,GHz.  The individual elements are arranged in a closely packed
hexagonal pattern.  The beam size at the 30-m telescope is $11''$ and
the spacing between the beams $20''$.  The observations were made in
March 1997 during a period of stable weather with zenith opacities
typically $<0.2$.  We monitored the sky opacity before and after each
subimage and mapped Mars every night to determine the absolute flux
scale.  To obtain a map, the object is scanned in azimuthal direction
including parts of blank sky on both sides to define a proper zero
level.  In addition, the sub-reflector of the telescope is oscillating
at a frequency of 2\,Hz which makes the beam switch between two
positions separated by $45''$ in the orientation of the scanning.
This yields an ``on-off'' measurement that cancels atmospheric
variations at short time scales.  The whole area of NGC\,4631 was
covered with a mosaic of 19 individual fields.  Their distribution is
rather uniform along the whole disk and, hence, the sensitivity drops
only at the outer edges of the final map.  Furthermore, we kept the
scanning orientation close to the minor axis of the galaxy by
carefully chosing the hour angles of the individual observations.
This minimizes spurious contributions and the noise in the map.  The
field presented here is about $15'\times 8'$ after the cutoff of the
edges with lower sensitivity.  In the central part, the noise is $\sim
2$ mJy/beam for the data smoothed to an angular resolution of $20''$
and rises to $\sim 3.5$ mJy/beam at the edges.

\subsection{NGC\,4631}
 
It is obviously best to choose edge-on galaxies for studies of weak
phenomena, because the lines of sight are long -- remember that the
dust emission in the mm regime is optically thin and thus the whole
disk contributes to the detectable flux. The 19-element bolometer made
it possible to map large areas, so we decided to re-observe
NGC\,4631. This moderately active galaxy has long been a favorite
candidate for an interacting system.  It is relatively nearby, at 7.5
megaparsecs (Mpc) (ref. 8; $1'$ corresponds to $\sim 2$\,kpc), and two
obvious companions are close by.  The dwarf elliptical NGC\,4627 is
situated $3'$ northwest of the nucleus and $30'$ to the southeast the
distorted spiral NGC\,4656 can be found.  The whole group has been
extensively studied in H{\sc i} (8,9) in order to understand the
traces of the interaction (cf.\ Fig.~\ref{fig:synops}).  According to
a modeling of the encounter (10) the prominent streamers of atomic gas
can be explained as being pulled out of the members of this group
during the interaction.  Presumably also as a result of the
interaction, the disk of NGC\,4631 has a disturbed appearance in the
optical continuum and H$\alpha$ line emission.  At two positions in
the disk highly energetic ``supershells'' have been found (11) of
which one is described as being caused by the impact of a high
velocity object (12).

Almost simultaneously with the early high-resolution H{\sc i}
observations a large radio halo was detected (13).  It is of
nonthermal origin and one of the most prominent radio halos known.
Detailed investigations (14) have subsequently shown that the magnetic
field lines of the galactic disk open into the halo -- in stark
contrast to most spiral galaxies, where the field is more or less
confined in the disk (15).  This magnetic field structure allows
electrons, cosmic ray particles and hot gas to escape from the active
disk into the halo.  ROSAT (Roentgen Satellite) detections (16,17)
show a large X-ray envelope that is a natural consequence of this
configuration.

The investigation of the molecular gas indicates rather normal
conditions, however.  The central region of $\sim 2'\times 1'$ was
completely mapped in the (1-0) and (2-1) transitions of CO (18).  In
addition, a major axis strip of $7'$ length was obtained with a higher
sensitivity of $\sim 20$\,mK. Some additional spectra at other
locations did not reveal significant emission.

\section{The distribution of the cold dust}
\subsection{The dusty disk}

\begin{figure}[ht]
\epsfig{bbllx=135,bblly=41,bburx=544,bbury=707,%
        figure=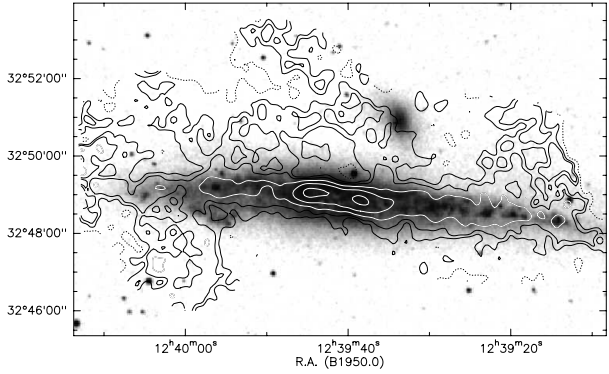,height=8.0cm,angle=270}   
\caption{Map of the $\lambda$ 1.2-mm emission of NGC\,4631, overlaid 
on an image taken from the Digital Sky Survey.  The levels are -6
(dotted), 6, 11, 21, 41, and 81 mJy/beam.  Only significant emission
is shown and the outer parts of the map with higher noise have been
cut off.  The small object north of the disk is the dwarf elliptical
galaxy NGC\,4627; the other companion, NGC\,4656, is situated about
half a degree away in the south-east.
}
\label{fig:map}
\end{figure}

\begin{figure}[ht]
\psfig{file=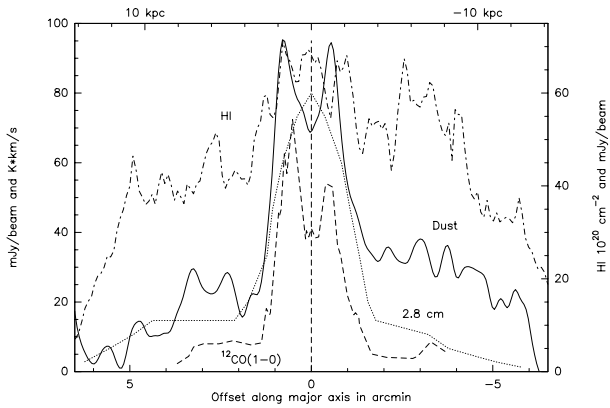,height=8cm,angle=270}
\vspace{2mm}
\caption{Cuts along the major axis; the CO and the 2.8\,cm data are 
provided by G.  Golla (14,18) and the H{\sc i} curve is extracted from
a map supplied by R. Rand (8).  The left axis gives the units for the
dust and CO emission, the right axis for H{\sc i} and cm continuum.
Similar to the galaxies NGC 4565 (6) and 5907 (7), the profiles of the
CO and the cm continuum drop off more steeply than that of the dust
and the H{\sc i}.  Note, however, that such cuts are one-dimensional
and miss emission that is close to, but not on the major axis.
}
\label{fig:macut}
\end{figure}

At a first glance, the $\lambda$ 1.2-mm emission of NGC\,4631
(Fig.\,\ref{fig:map}) is characterized by a narrow, extended disk,
with a double-peaked central region.  About three times weaker than
the brightest peaks, at a level of $\simeq 30$\,mJy/beam, the disk is
stretched out to a distance of $\sim 13$\,kpc on either side of the
nucleus, gently decreasing in the west, and at a relatively constant
level for some 10\,kpc in the east -- see Fig.\,\ref{fig:macut}.  This
distribution is similar to that found in NGC\,4565 (6): The
correlation between dust and CO is restricted to the nuclear region,
whereas the dust in the outer parts of the disk seems to follow the
H{\sc i}.  The CO emission drops in a similarly steep manner as the
centimetric radio continuum: At a radius of $2.5'$ ($=5.5\,$kpc) it
has decreased to a tenth of the peak value, and a bit further out, no
CO emission could be detected at all.  At this radius, the dust
emission is still at about one-third of the peak level, however, and
it actually stretches out at least to the edge of the optical disk.
In fact, the limits are set by the border of the map, not by the
vanishing emission.  The old bolometer map (5) is limited to the
innermost part due to its restricted coverage and sensitivity.

The narrow main emission ridge is somewhat surprising given the fact
that NGC\,4631 does not even show a dust lane.  Its thickness is
similar to that of the undisturbed edge-on galaxies NGC\,4565 (6) or
NGC\,891 (3).  This suggests that the dust is concentrated in the
midplane of a galaxy in any case (cf.  ref.\ 19), and the absence of
an optical dust lane may just reflect a high clumpiness with large
``holes''.

\begin{figure}[ht]
\epsfig{bbllx=281,bblly=41,bburx=544,bbury=707,%
        figure=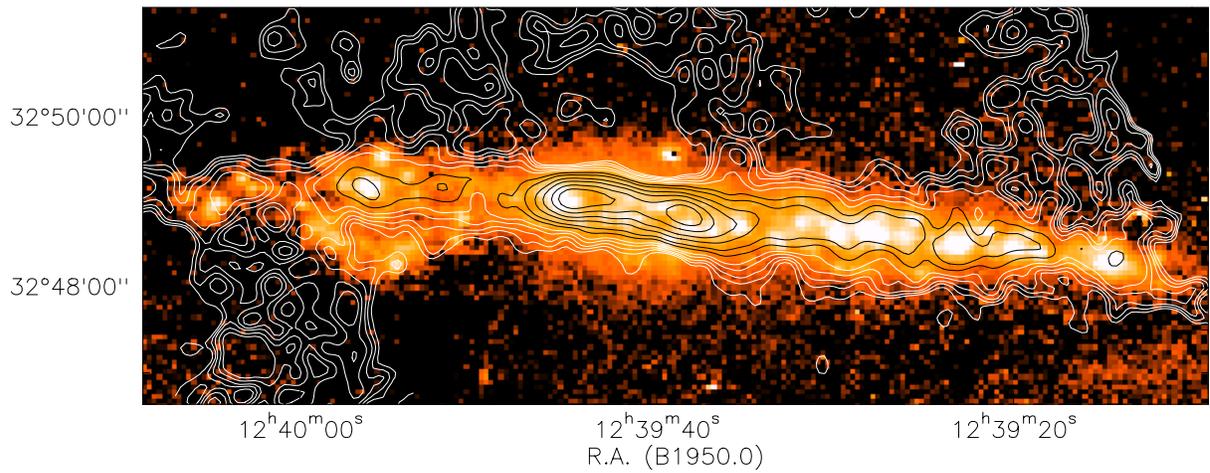,height=16.0cm,angle=270}   
\caption{Comparison of the $\lambda$ 1.2-mm emission of NGC\,4631 
with the H$\alpha$ emission which has been smoothed to $6''$.  The 
dwarf elliptical galaxy NGC\,4627 is invisible here.  Note the large 
disturbed area in the west with its subsequent `cutoff'.  Near the 
center, the dust and H$\alpha$ tend to be anti-coincident whereas in 
the outer disk there are clear correspondences.
} 
\label{fig:hamap}
\end{figure}

\begin{figure}[ht]
\psfig{file=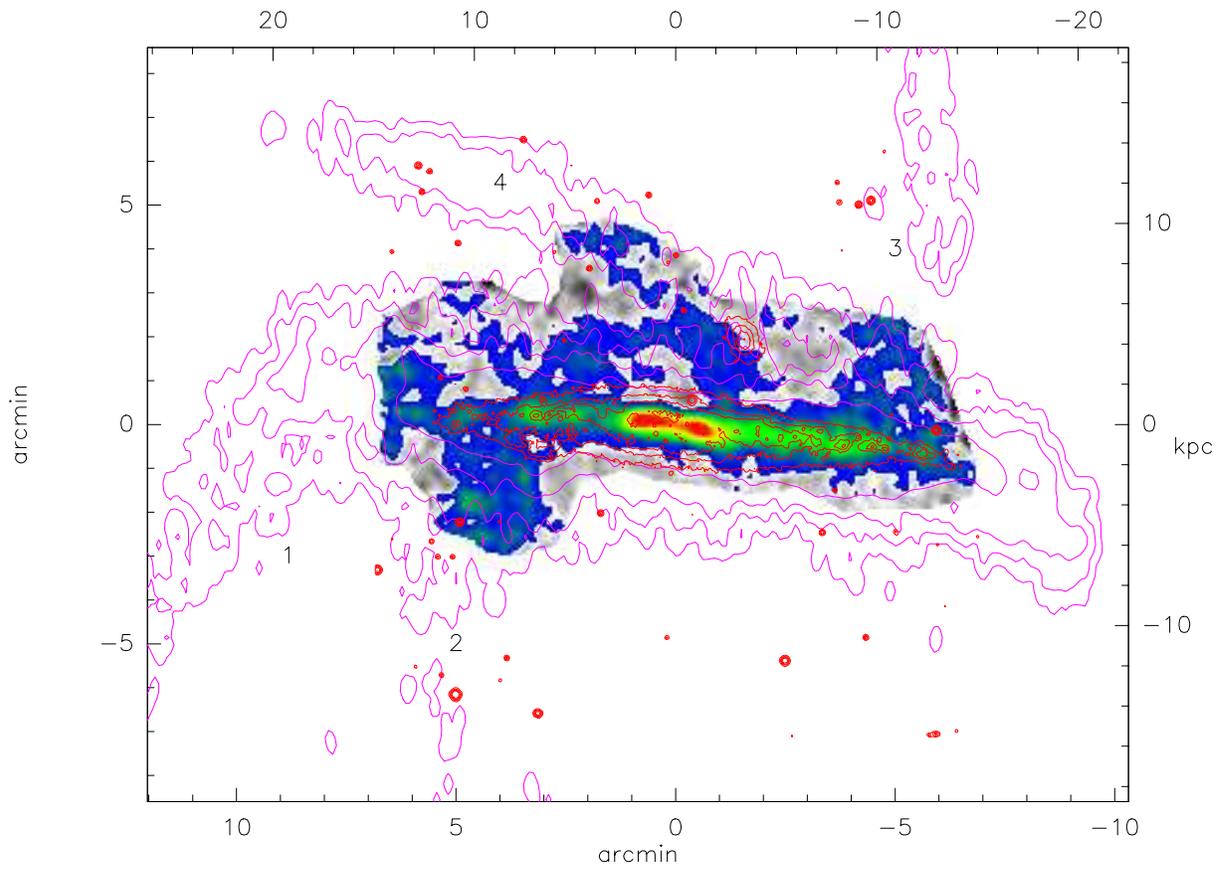,height=16cm,angle=270}
\caption{An overview of the distribution of the extraplanar H{\sc i} 
(purple contours) and cold dust (in color); the H{\sc i} is smoothed 
to $22''$ spatial resolution.  For comparison, optical isophotes are 
added in red.  The H{\sc i} spurs are numbered as introduced in ref.\
9. The axis labels are given in arc minutes from the center of
NGC\,4631 on the lower and left sides; the other sides gives the
projected length scale for an assumed distance of 7.5\,Mpc.
}
\label{fig:synops}
\end{figure}

A closer inspection shows some radial variations in comparison with 
the H$\alpha$ map (Fig.~\ref{fig:hamap}).  Near the center, the 
optically bright regions surround the dust emission peaks.  Here, the 
bulk of the optical emission is produced in the central region and 
later absorbed on its way through the disk.  In contrast, in the outer 
parts of the disk several dust maxima coincide with optically bright 
spots.  It is unlikely that all of these places are just accidental 
line-of-sight coincidences.  The source of the radiation could be 
young stars in their dusty birthplaces (e.g.\ in a spiral arm 
tangent).  In any case, the material along the light path should be 
rather transparent.  Such large variations of the opacity are not 
surprising for spiral galaxies, however (see ref.\ 20 
for a compilation).

\subsection{The extraplanar dust}

Completely new is the detection of significant dust emission out to
$z$-distances of at least 10\,kpc.  The distribution of this
intergalactic dust strongly suggests that it has been brought there by
the same mechanism that formed the four H{\sc i} streamers (8,9). All
three H{\sc i} spurs that are touched by our map are connected with
the thin optical disk by corresponding dust features (see
Fig.~\ref{fig:synops}).  The disk formed by the atomic gas is much
thicker than the optical or dust emission disk, however, so that most
of the $\lambda$ 1.2-]mm emission still lies within their boundaries.
On the other hand, CO emission could only be detected out to $z\sim
1\,$kpc.  So either there is hidden molecular gas in those outer
regions or the dust is associated with atomic gas.

Although these streamers follow the already known H{\sc i} features
rather closely, north of the center of NGC\,4631 a structure has been
unveiled that was invisible in the thick atomic gas disk: A giant arc
spans over the central region with its footpoints about 4\,kpc east
and west of the central region.  The eastern footpoint is situated
opposite the onset of the southern streamer 2, and the western part of
the arc is blending into spur 4.  The thickness of the disk emission
seems to be reduced in the central region, but this might be an
artifact of the data-reducing technique.  The question arises
regarding whether there is a common origin for these extraplanar
structures.

\section{Origin of the extraplanar dust}

Unfortunately, we do not have any velocity information from the
$\lambda$ 1.2-mm continuum observations, and hence have to rely on
indirect arguments for the determination of the history of this dust
distribution.  To some degree, it seems reasonable to assume that the
H{\sc i} velocities are a good indicator also for the dust kinematics.
Indeed, there are at least two clear coincidences between H{\sc i}
velocity anomalies and the arc.  The velocity gradient along the major
axis is much steeper in the northern part of the disk than in the
southern.  A possible interpretation suggests that the disk is
somewhat inclined and warped along the line of sight, so we would be
looking at different portions of the galaxy (8) -- north of the center
at the inner disk with a steeper gradient and south of it at the near
part of the outer disk.

\begin{figure}[ht]
\epsfig{bbllx=143,bblly=41,bburx=544,bbury=753,%
        file=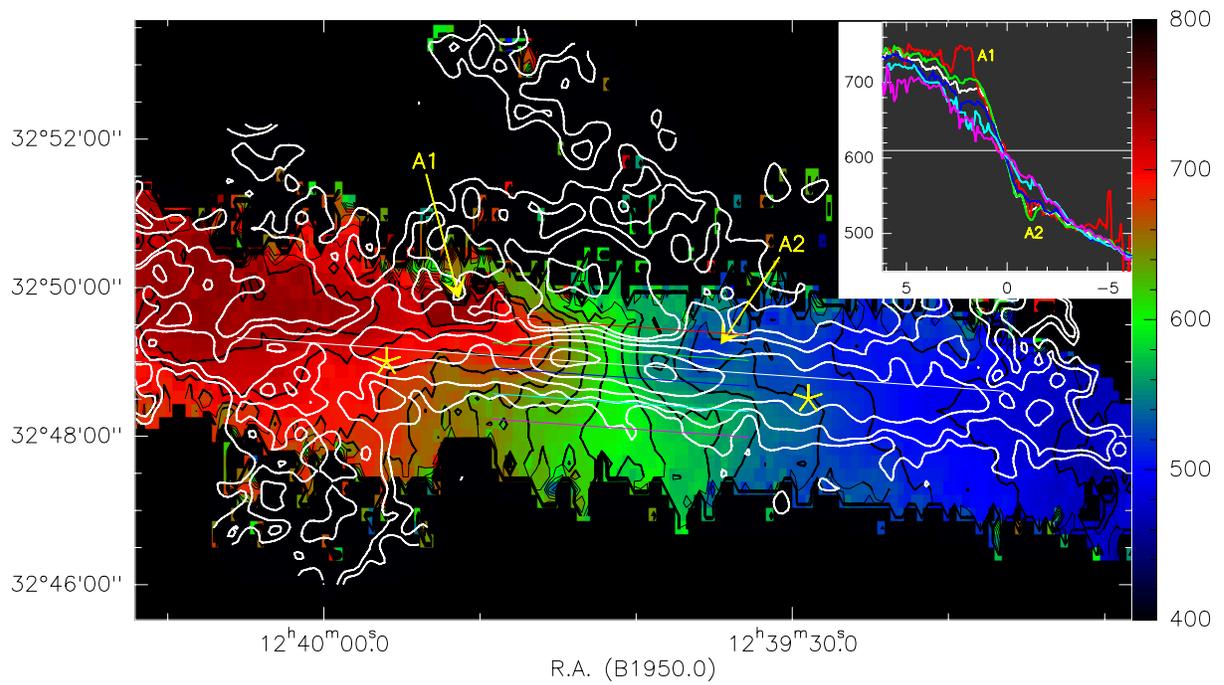,height=16cm,angle=270}
\caption{A comparison between the cold dust emission (at $20''$ 
resolution) and the velocity field of the atomic gas (at $12''\times
22''$).  Velocity contours are plotted every 20\,km/s (cf.\ 
also the right-hand scale).  Note the velocity anomalies at the foot 
points of the arc (arrows).  The insert shows position-velocity curves 
parallel to the major axis; the box is labelled in velocity vs.\ 
minutes of arc.  The red line goes through A1, the green line through 
A2 as indicated by short lines in the main plot.  In the southern disk 
the gradient is flatter (purple line).  The stars mark the position of 
the two H{\sc i} supershells.
}
\label{fig:rotbol}
\end{figure}

If we compare the velocity field with the dust emission (see 
Fig.~\ref{fig:rotbol}), a different explanation arises.  In addition 
to the steeper gradient, the northern disk shows anomalous velocity 
components at two places -- and they coincide perfectly with the 
footpoints of the arc.  These anomalous components are visible as 
regions with almost closed isovelocity contours in 
Fig.~\ref{fig:rotbol} (marked as A1 and A2).  Although we are not able 
to determine the precise location of the extragalactic dust with 
respect to the disk, it seems clear that there is a local interaction.  
The additional velocity components point towards us in the west and 
away from us in the east.

In addition there are two structural indications, both of them
suggesting that material has followed the arc in an counterclockwise
direction and hit the disk in the east: ({\it i}) The appearance of
the disk in the light of the H$\alpha$ line is much more disturbed at
the eastern footpoint of the arc (cf.\ Fig.~\ref{fig:hamap}: In
broadband red light the eastern part seems truncated.  The disturbed
region lies between the arc and the streamer 2, whereas the western
side seems unaffected by the arc or streamers 3 and 4.  ({\it ii}) In
the H{\sc i} emission of NGC\,4631 two supershells have been found
near the midplane and subsequently have been modeled.  Shell~2 (west
of the nucleus at Right Ascension 12$^h$39$^m$30$^s$) seems to be an
expanding bubble, whereas shell~1 (at 12$^h$39$^m$56$^s$) is most
probably caused by an impact of a cloud with a mass of $\sim 10^{7}
M_{\odot}$ at a speed of 200 km/s coming {\it from the north} (12).
The geometry of the collision is very well constrained by its
kinematical signature in the H{\sc i} data and indicates that he
material not only came from the north but, moreover, must have had a
velocity component away from us.  The location of this shell is
between arc and spur 2.  This fits to the other indications of the
trajectory of the arc (the shells and the anomalous components
mentioned above are distinct features).

If we try to describe all extraplanar dust features as a single trail,
its path could be as follows: starting from spur 3, it runs through
the western disk, follows it to the western footpoint of the arc,
sweeps along the arc, penetrates the eastern disk and leaves it via
spur 2.  In the way, some material is forced away: for example, to
follow streamer 4.  Of course, such a concatenation is purely
speculative and, at this point, merely meant to summarize the
structure of the dust distribution.  In principle, the concept of a
continuous trail is a natural explanation within the framework of an
interaction, however.  In any case, we have to answer the question
regarding which of the three involved galaxies has left the dust
traces.

The interaction has been modeled rather early on the basis of the
first high-resolution H{\sc i} data (10).  This description suggests
that the bridge 1 and the streamer 4 had been parts of the disk of
NGC\,4631.  The two perpendicular features 2 and 3 are made from
material pulled out of the now dwarf elliptical NGC\,4627 which might
have been of a different type before.  However, the H{\sc i} emission
of structures 2 and 3 is only weakly connected with the disk of
NGC\,4631 and not at all with the dwarf.  It has to be stressed that
this model is by no means unique -- three involved galaxies open a
vast parameter space for an interaction scenario.

Today, the location of part of the H{\sc i} gas can be determined even
in the third dimension with the help of X-ray data: The soft band is
very sensitive to absorption by atomic hydrogen, and it can be safely
concluded from the distributions that the southern part of the disk is
at the near side (16,17).  Looking a bit more in detail, the onset of
spur 4 seems to be located in front of the X-ray halo (cf.\ Figure 6
of ref.\ 17) but, higher up, the X-ray emission becomes stronger
again, suggesting that this streamer is pointing away from us.  Such
additional information will help to constrain the parameters of the
interaction much better than it was possible before.

\section{Discussion and outlook}
\subsection{Origin of the mm emission}

NGC\,4631 is characterized not only by the interaction with two other 
galaxies but also by a unusually large radio halo.  In particular the 
enhanced star formation might be responsible for the uncommon magnetic 
field configuration and thus for the radio and X-ray halo -- could 
this possibly produce extraordinary continuum emission at $\lambda$ 
1.2\,mm as well?

In addition to radiation from dust, there are three candidates for the
source of radiation: free-free emission in ionized clouds, nonthermal
emission and line radiation within the bandpass of the detector.  The
free-free emission should be correlated with H$\alpha$ -- so a
contribution is only expected in the disk since the sparse high-$z$
H$\alpha$ emission (21) is not correlated with dust features.  Within
the disk, even in the brightest spots the emission measure reaches
only a relatively low value of 1000\,pc$\cdot$cm$^{-6}$ (22).  So the
{\it total} flux due to free-free radiation is of the order of 10\,mJy
only.

The synchrotron emission is very extended around NGC\,4631 at cm
wavelengths (14).  From the fluxes and the determined spectral index
(23) we can estimate its contribution in our mm band.  In the center,
the spectral index is rather flat at about $\alpha\sim -0.65$.  The
peak flux here is 60\,mJy/$84''$\,beam at 10.55\,GHz, this translates
into 0.3 mJy/$20''$\,beam at 230\,GHz.  Outside the disk the spectral
index is even steeper, so this contribution is smaller than a tenth of
the noise level.  The only significant addition thus comes from the
molecular line emission: The eastern peak reaches 70
K$\cdot$km\,s$^{-1}$ of $^{12}$CO(2-1) emission.  Within our $20''$
beam, this contributes to a flux density of $\sim 20$\,mJy/beam (cf.\
ref.\ 4), about a fifth of the total value.  In the outer disk,
however, the line emission adds $<3$\,mJy/beam -- just the figure of
the noise level in the map.
 
So, the observed mm continuum radiation is pure thermal dust emission
almost everywhere, but how did the dust reach such enormous
$z$-heights?  Could it simply be blown out of the active plane by
large-scale winds?  Outflows are known for many galaxies with strong
starbursts like, for example, M\,82 (see, e.g., refs.\ 24 and 25).
But the disk of NGC\,4631 is forming stars at an only moderately
enhanced rate (18).  H$\alpha$ kinematic data as a more direct tool do
not show signs of a gas outflow from the disk (22) -- in contrast to
M\,82 which has a similar orientation (24).  Peculiar velocities have
been found, but they are more likely explained by the direct influence
of the interaction.

\subsection{Temperatures and Masses}

The dust around the disk of NGC\,4631 seems to be cold -- the
spatially resolved IRAS CPC data as an indicator for warmer dust show
at most a somewhat thicker disk near the center (26).  Its width is of
the order of $90''$ at $100\mu$m and $45''$ at $50\mu$m wavelength
using the published beam sizes of $95''$ and $80''$, respectively.  No
significant emission can be seen further out.  Unfortunately, the CPC
instrument could not be calibrated properly, and because of the
uncertainties of the order of $\pm60\% $ we could not derive dust
temperatures from these data.

In view of the moderate star forming activity the presence of very
warm material does not seem very likely.  Presumably the temperature
of the coldest dust component is rather low even in the disk --
similarly to other spirals in which it could be measured so far (2,6).
We expect, therefore, temperatures in the range of 15 to 20 K for the
disk, and even lower values for the extraplanar dust.  To determine
such temperatures, observations in the sub-mm range between 1\,mm and
100\,$\mu$m are needed, but, here, the opaque atmosphere renders them
very difficult.  This is why only few data of spiral galaxies have
been published in this range (e.g.\ refs.\ 2 and 27).  We will do more
observations at $\lambda\lambda$ 450\,$\mu$m and 850\,$\mu$m to
complement the existing data.  Together, they should provide the
spectral information needed for the determination of the temperatures.

Until then, we have to postpone the proper determination of masses and
energies from the dust emission.  In principle, the dust mass is
directly proportional to the detected flux $S_{\lambda}$.  With some
additional assumptions about the dust and gas properties, we can
derive the hydrogen column density.  In the millimeter/sub-millimeter
regime, the observed flux density per beam (of half power beam width
$\theta$) produced by dust of temperature $T_{d}$ is given by
$S_{\lambda}=
1.13\,\theta^{2}\,(1-e^{\tau_{\lambda}})\,B(\lambda,T_{d})$.  Here,
$\tau_{\lambda}=\sigma_{\lambda}^{\rm H}\cdot N_{\rm H}$ gives the
dust absorption cross section per hydrogen atom, and
$B(\lambda,T_{d})$ is the Planck law for the radiation of a black body
(see ref.\ 1 for a derivation, and ref.\ 6 for a determination of a
possibly typical value).  Such a calculation yields about
$2\cdot10^{9}$ M$_{\odot}$ of hydrogen for the gas associated with a
dust component of 21.5\,K in the central region (5).  As already
stated above, the warmer dust radiates much more efficiently, and,
hence, a second dust component at 55\,K is associated with $<1$\% of
this mass (5).

These values are poorly constrained, however, and we face large 
uncertainties in the temperatures of the coldest component.  If we 
nevertheless assume 15\,K for the extraplanar dust the total gas mass 
in the arc should be of the order of $1.7\cdot10^{9}$ M$_{\odot}$, 
somewhat less than the mass in the central region.  A similar value is 
found for its presumed continuation, spur 2, so that the total gas 
mass in this structure would be about half of the atomic hydrogen mass 
of the whole galaxy (8).

\subsection{Conclusion}

Obviously, the interaction is the dominant event in the history of
NGC\,4631.  It has caused a partial disruption of the disk, triggered
the star formation activity, provoked an upturned magnetic field
configuration, and, last but not least, left trails of atomic gas and
dust all around the disk.  It is therefore crucial to know exactly how
the interaction took place.  The existing model (10) was obtained by
using a ``trial-and-error'' method for the parameters. It is virtually
impossible to find the proper solution that way; the number of
possible parameter sets for an interaction in such a group with three
members is simply too large.  Moreover, the now existing data
(including the distribution of the cold dust) give new and more
detailed information to check the model against.  The problem is how
to explore the parameter space to find the best set.  Now there are
new, powerful techniques available like so-called `genetic codes'
(28), so we hope to obtain a much better view of the interaction,
which in turn should help to fix the roots of the related phenomena.

In any case, the detection of cold dust outside of a galactic disk has 
unveiled the existence of hitherto invisible baryonic matter in the 
halo region.  This will not solve the dark matter problem, even using 
a favorable estimation of the associated mass, but it is a step 
towards a more ``normal'' neighborhood of galaxies -- containing less 
exotic material, probably simpler to understand, but in any case 
easier to investigate.

\vspace{2em}\noindent
{\small {\bf Acknowledgments} We thank G. Golla for the H$\alpha$
picture and the CO data set, R. Rand for several sets of H{\sc i}
data, A. Vogler for the ROSAT data and J. Kerp for help with their
interpretation. F. Combes supplied the trajectories for her model so
that we got an idea of the time evolution.  Part of this work was
supported by the Deutsche Forschungsgemeinschaft within the frame of
SFB301.}

\small
\noindent

\subsubsection*{References}

1. Cox, P. \& Mezger, P.G. (1989)
  {\it Astron.\ Astrophys. Rev} {\bf 1}, 49-83 \\
2. Kr\"ugel, E., Siebenmorgen, R., Zota, V. \& Chini, R. (1998),
  {\it Astron.\ Astrophys.} {\bf 331}, L9-L12 \\
3. Gu\'elin, M., Zylka, R., Mezger, P.G., Haslam, C.G.T., Kreysa, E., Lemke, 
  R. \& Sievers, A. (1993) {\it Astron.\ Astrophys.} {\bf 279}, L37-L40 \\
4. Gu\'elin, M., Zylka, R., Mezger, P.G., Haslam, C.G.T. \& Kreysa, E.
  (1995) {\it Astron.\ Astrophys.} {\bf 298}, L29-L32 \\
5. Braine, J., Kr\"ugel, E., Sievers, A. \& Wielebinski, R. (1995) 
  {\it Astron.\ Astrophys.} {\bf 295}, L55-L58 \\
6. Neininger, N., Gu\'elin, M., Garc\'{\i}a-Burillo, S., Zylka, R. \& 
  Wielebinski, R. (1996) {\it Astron.\ Astrophys.} {\bf 310}, 725-736 \\
7. Dumke, M., Braine, J., Krause, M., Zylka, R., Wielebinski, R. \& 
  Gu\'elin, M. (1997) {\it Astron.\ Astrophys.} {\bf 325}, 124-134 \\
8. Rand, R.J. (1994) {\it Astron.\ Astrophys.} {\bf 285}, 833-856 \\
9. Weliachew, L., Sancisi, R. \& Gu\'elin, M. (1978) 
  {\it Astron.\ Astrophys.} {\bf 65}, 37-45 \\
10. Combes, F. (1978) {\it  Astron.\ Astrophys.} {\bf 65}, 47-55 \\
11. Rand, R.J. \& van der Hulst, J.M. (1993) {\it  Astron.\ J.} {\bf 105},
  2098-2106 \\
12. Rand, R.J. \& Stone, J.M. (1996) {\it  Astron.\ J.} {\bf 111}, 190-196 \\
13. Ekers, R.D. \& Sancisi, R. (1977) 
  {\it Astron.\ Astrophys.} {\bf 54}, 973-974 \\
14. Golla, G. \& Hummel, E. (1994) 
   {\it  Astron.\ Astrophys.} {\bf 284}, 777-792 \\
15. Dumke, M., Krause, M., Wielebinski, R. \& Klein, U. (1995) 
  {\it Astron.\ Astrophys.} {\bf 302}, 691-703 \\
16. Wang, Q.D., Walterbos, R.A.M., Steakley, M.F., Norman, C.A. \&
  Braun, R. (1995) {\it Astrophys.\ J.} {\bf 439}, 176-184 \\
17. Vogler, A. \& Pietsch, W. (1996) 
  {\it Astron.\ Astrophys.} {\bf 311}, 35-48 \\
18. Golla, G. \& Wielebinski, R. (1994) 
  {\it Astron.\ Astrophys.} {\bf 286}, 733-747 \\
19. Xilouris, E.M., Byun, Y.I., Kylafis, N.D., Paleologou, E.V. \& 
  Papamastorakis, J. (1999)  
  {\it Astron.\ Astrophys.}, {\bf 344}, 868-878 \\
20. Davies, J.I. \& Burstein, D. (eds.), (1995) 
  {\it The Opacity of Spiral Disks} (Kluwer, Dordrecht), Vol.\ 469 \\
21. Rand, R.J., Kulkarni, S.R. \& Hester, J.J. (1992)
  {\it Astrophys.\ J.} {\bf 396}, 97-103 \\
22. Golla, G., Dettmar, R.-J. \& Domg\"orgen H. (1996) 
  {\it Astron.\ Astrophys.} {\bf 313}, 439-447 \\
23. Golla, G. (1993) {PhD Thesis}, University of Bonn \\
24. McKeith, C.D., Greve, A., Downes, D. \& Prada, F. (1995) 
  {\it Astron.\ Astrophys.} {\bf 293}, 703-709 \\
25. Bregman, J.N., Schulman, E. \& Tomisaka K. (1995) 
  {\it Astrophys.\ J.} {\bf 439}, 155-162 \\
26. van Driel, W., de Graauw, Th., de Jong, T. \& Wesselius, P.R. (1993) 
  {\it  Astron.\ Astrophys.\ Sup.} {\bf 101}, 207-252 \\
27. Alton, P.B., Bianchi, S., Rand, R.J., Xilouris, E.M., Davies,
J.I. \& Trewhella, M. {\it Astrophys.\ J.} {\bf 507}, L125-L129 \\
28. Theis, Ch. (1999) {\it Reviews in Modern Astronomy} {\bf 12} (in press)
%

\normalsize

\end{document}